\documentclass[12pt]{article}

\usepackage{geometry}
\usepackage{graphicx} 
\usepackage{amsmath, amssymb}
\usepackage{natbib}
\usepackage{booktabs}
\usepackage{multirow}
\usepackage{sectsty} 
\usepackage{helvet}
\usepackage{enumerate}
\usepackage{url}
\usepackage{xcolor}
\usepackage{subfigure}

\newcommand{\cowscape}{Cow\-Scape}
\newcommand{\hide}[1]{{}}

\usepackage[helvet]{sfmath}

\newcommand{\MPI}{Max Planck Institute for Multidisciplinary Sciences, 37077 G\"ottingen, Germany}
\newcommand{\MIA}{Microscopic Image Analysis Group, Jena University Hospital, 07743 Jena, Germany}
\newcommand{\IMP}{Institute for Molecular Pathology, Vienna, Austria}

\newcommand{\POSMPI}{1}
\newcommand{\POSMIA}{2}
\newcommand{\POSIMP}{3}

\title{\cowscape: Quantitative reconstruction of the conformational landscape of biological macromolecules from cryo-EM data}
\author{%
  Felix Lambrecht\,$^{\text{\sf \POSMPI}}$, 
  Andreas Kr\"opelin\,$^{\text{\sf \POSMIA}}$,
  Mario L\"uttich\,$^{\text{\sf \POSMPI}}$,
  Michael Habeck\,$^{\text{\sf \POSMIA,\POSMPI,*}}$, \\
  David Haselbach\,$^{\text{\sf \POSMPI,\POSIMP,*}}$, 
  Holger Stark\,$^{\text{\sf \POSMPI,*}}$
}
\newcommand{\myaffiliation}{%
  \noindent $^{\text{\sf 1}}$\MPI\\
  $^{\text{\sf 2}}$\MIA\\
  $^{\text{\sf 3}}$\IMP\\\\
}
\newcommand{\myemail}{*E-Mail: hstark1@gwdg.de; david.haselbach@imp.ac.at; michael.habeck@uni-jena.de}

\begin{document}

\maketitle
\myaffiliation
\myemail

\abstract{
  \noindent Cryo-EM data processing typically focuses on the structure of the main conformational state under investigation and discards images that belong to other states. 
  This approach can reach atomic resolution, but ignores vast amounts of valuable information about the underlying conformational ensemble and its dynamics.
  \cowscape{} analyzes an entire cryo-EM dataset and thereby obtains a quantitative description of structural variability of macromolecular complexes that represents the biochemically relevant conformational space.
  By combining extensive image classification with principal component analysis (PCA) of the classified 3D volumes and kernel density estimation, \cowscape{} can be used as a quantitative tool to analyze this variability.
  PCA projects all 3D structures along the major modes spanning a low-dimensional space that captures a large portion of structural variability.
  The number of particle images in a given state can be used to calculate an energy landscape based on kernel density estimation and Boltzmann inversion. 
  By revealing allosteric interactions in macromolecular complexes, \cowscape{} allows us to distinguish and interpret dynamic changes in macromolecular complexes during function and regulation.
}

\clearpage

\section{Introduction}
Most biological processes in the cell are not driven by individual biomolecules but rather by large macromolecular assemblies comprising numerous proteins and/or nucleic acids \citep{Gavin2002}.
These complexes can be considered as molecular machines: they tend to be highly dynamic and undergo significant conformational rearrangements during their reaction cycles.
In their natural biological context, molecular machines are regulated, for example, by the binding of external factors that restrict the conformational freedom in a functionally relevant manner. 
Understanding the function of macromolecular machines requires that we consider the entire conformational landscape that the macromolecular complex explores during regulation and its reaction cycle \citep{Fischer2010, Haselbach2017, Haselbach2018, Singh2020}. 

The current focus of single-particle cryo-EM is mainly on high-resolution structure determination of a single static state. 
Realizing the potential of single-particle cryo-EM for studying both the structure and dynamics of macromolecular machines, necessitates that we go beyond the current state-of-the-art in cryo-EM data processing. 
The most complete description would be a high-resolution molecular ``movie'' of the entire accessible conformational space.
Such a movie would, for example, allow us to better understand allosteric regulation effects resulting from the binding of a ligand (such as a drug) that affects the activity of the complex.

Historically, X-ray crystallography has dominated as the method for determining high-resolution structure of macromolecular complexes.
Yet X-ray crystallography is mainly limited to static structure determination, because it studies copies of the molecule in roughly the same conformation packed into a crystal lattice.
Cryo-EM, in contrast, images individual molecules in a frozen hydrated state with no bias to adopt a conformation preferred by crystal packing.
The range of structural heterogeneity due to conformational dynamics is therefore significantly higher in cryo-EM than in X-ray crystallography.
This can be very problematic for high-resolution structure determination with cryo-EM and is reflected by the current statistics of the Electron Microscopy Data Bank (EMDB) that reveal how detrimental structural heterogeneity can be.
Despite ground-breaking technical developments that now allow the determination of single-state structures up to atomic resolution, the average resolution of cryo-EM structures deposited in EMDB \citep{EMDB2023}, is significantly lower. 
Indeed, any mixture of conformations (due to either low sample quality or the presence of multiple conformations) will always be resolution-limiting in cryo-EM.

Some aspects of this heterogeneity can be tackled by using computational tools to sort cryo-EM images according to differences in composition and conformation.
A large dataset of cryo-EM images can thus be purified {\em in silico} to compensate for a higher level of conformational and structural heterogeneity.
Computational tools have been developed for this computational sorting, including two-dimensional (2D) PCA \citep{Elad2008}, maximum likelihood-based classification in 2D/3D \citep{Scheres2016, Moriya2017}, and manifold embedding of raw images \citep{Dashti2014}.
However, while these computational tools effectively distinguish between distinct conformational states of the molecules, their ability to separate continuous conformational differences is usually limited.

Currently, the most widely used classifications in cryo-EM are based on maximum-likelihood \citep{Sigworth1998}.
They do not require prior knowledge and are easy to use via powerful software packages such as Relion \citep{Scheres2016} or cryoSparc \citep{Punjani2017}. 
A cryo-EM project typically aims to computationally purify particle images that belong to the main conformational state present in the dataset.
Several rounds of classification remove particle images that do not match the main conformational state. 
The ``purified'' sub-population can then be refined to high resolution.
In some cases, this approach discards more than 90\%{} of all images to reach a level of homogeneity that allows the calculation of a high-resolution structure.
Thus, the majority of images does not belong to the best sub-population in many cryo-EM datasets. 
The discarded data are not analyzed further, such that it remains unclear why the majority of images do not contribute to the high-resolution structure.
Nonetheless, the discarded data may still contain interesting and relevant information.

In an ideal situation, the entire data would be used to computationally purify all well-defined states that exist in a given dataset, rather than only the most evident ones, and each image would be classified into its corresponding three-dimensional (3D) structure.
However, such an approach would require massive image statistics, because the conformational freedom of the entire  macromolecular complex should be computationally sorted to understand its full dynamics.
In recent years, a number of computational tools have been developed to tackle conformational heterogeneity in cryo-EM datasets \citep{Tang2023, Toader2023}.
These range from manifold embedding \citep{Dashti2014, Dashti2020}, to multi-body refinement \citep{Nakane2018} and machine-learning approaches \citep{Chen2021, Zhong2021}, to name just a few. 

The first attempt to determine the entire conformational landscape from a large image dataset was the exhaustive computational analysis of an E. coli 70S ribosome trapped in intermediate states of tRNA retro-translocation \citep{Fischer2010}.
In this study, two million particle images were used to separate 50 conformational states by a hierarchical computational purification procedure as well as by visual inspection of the calculated 3D structures based on the tRNA positions on the ribosome.
Following the tRNA motions by visual comparison of all structures allowed the order of the structures to be determined along a conformational trajectory.
After taking the particle statistics into account, the first energy landscape based on cryo-EM data was determined \citep{Fischer2010}. 
The entire procedure required extensive manual image processing, as well as prior knowledge about tRNA motions, to find the order of all states which was later confirmed by molecular dynamic simulations \citep{Bock2013}.
It is therefore impractical to apply this approach to samples for which little or no prior knowledge is available.

\begin{figure}[h!]
  \centering\includegraphics[width=0.9\textwidth]{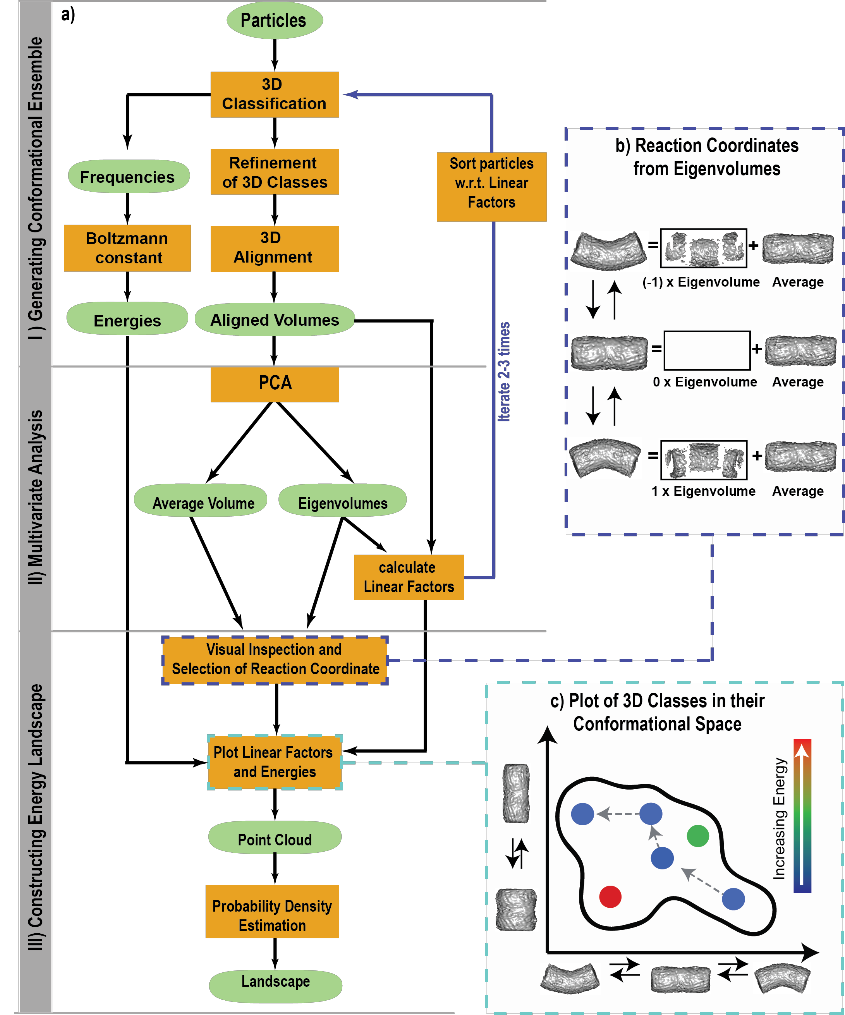}
  \caption{\textbf{The \cowscape{} algorithm.} a) Overview of the algorithm. \cowscape{} is roughly divided into three steps: I) generate the conformational ensemble by applying classification algorithms; II) carry out PCA; and III) make a final selection of reaction coordinates and construct the energy landscape. b) Interpretation of eigenvolumes. Each volume of a trajectory (left column) can be described as a linear combination of the eigenvolume multiplied with a linear factor and the average volume. Hence, each eigenvolume describes a conformational motion. c) Using the linear factors, volumes can be plotted into a 2D conformational space with the respective energies (indicated by a colour gradient)}\label{fig:cowscape}
\end{figure}

Despite the recent progress in addressing continuous conformational hetereogeneity in cryo-EM datasets \citep{Tang2023, Toader2023}, we prefer a conceptually simple and interpretable approach that does not require exhaustive training, but can be applied directly to any dataset. 
To realize this idea, we have developed \cowscape{}, a computational approach to obtain a quantitative description of conformational variability of macromolecular complexes (Fig. \ref{fig:cowscape}a).
\cowscape{} combines extensive maximum-likelihood 3D classification and PCA (3D-PCA) to determine all possible conformational states without the need for any prior knowledge about the complexes.
PCA is used in several approaches to analyze heterogeneity in cryo-EM (e.g. by \citet{Tagare2015} as well as \citet{Punjani2021}).
Despite its limitations in resolution \citep{Sorzano2021}, we find that 3D-PCA is a highly useful tool to study large-scale conformational changes, in particular when combined with biochemical knowledge. 
\cowscape{} analyzes the similarities and differences of the states in a meaningful and automated manner and then orders the states by a PCA of all calculated and aligned 3D volumes.
The number of states can range from several tens to several hundreds of structures.
The main modes of conformational variations are directly obtained by the major PCA eigenvectors (Fig. \ref{fig:cowscape}b).
Taking two major modes allows the conformational landscape to be plotted, which provides a plausible trajectory of conformational changes (Fig. \ref{fig:cowscape}c).
Furthermore, the number of particle images used to calculate the structures of each conformational state is known.
Based on the particle numbers, \cowscape{} can convert the conformational landscape into an energy landscape by applying Boltzmann's law.
We have applied \cowscape{} to large macromolecular complexes including the human 26S proteasome \citep{Haselbach2017}, the human B-act spliceosome \citep{Haselbach2018} and the fatty acid synthase \citep{Singh2020}.
%
%
In all cases, \cowscape{} allowed us to determine structures of unknown conformations and obtain new structural insight into functionally relevant regulatory mechanisms.
Here, we present the details of how \cowscape{} works and a demonstration of its general applicability. 

\section{Results}
The \cowscape{} algorithm is available in the COW image processing suite (\url{www.cow-em.de}).
\cowscape{} provides tools for 3D classification and refinement, the generation of conformational ensembles and energy landscapes, for plotting data as a 2D heat map, and for visualizing these data as a 3D landscape.
The first step is an extensive 3D classification of all data identified as true particle images within a cryo-EM dataset (for an overview see Fig. \ref{fig:cowscape}).
3D classification can be performed with various image processing packages \citep{Scheres2016, Moriya2017, Grant2018, Punjani2017} including the COW package.
After classification, each sub-population is refined to the highest possible resolution.
Notably, 3D classification provides both an overview of the conformational ensemble and the number of particles assigned to each class.
These numbers encode highly valuable quantitative information, because a thermodynamically stable conformation will be observed with a higher particle frequency than a thermodynamically less-favored state.
The particle number per conformational sub-population thus directly translates into free energy differences and can be used to construct an energy landscape via Boltzmann inversion (see Supplementary Methods).

\begin{figure}[h!]
  \centering\includegraphics[width=0.9\textwidth]{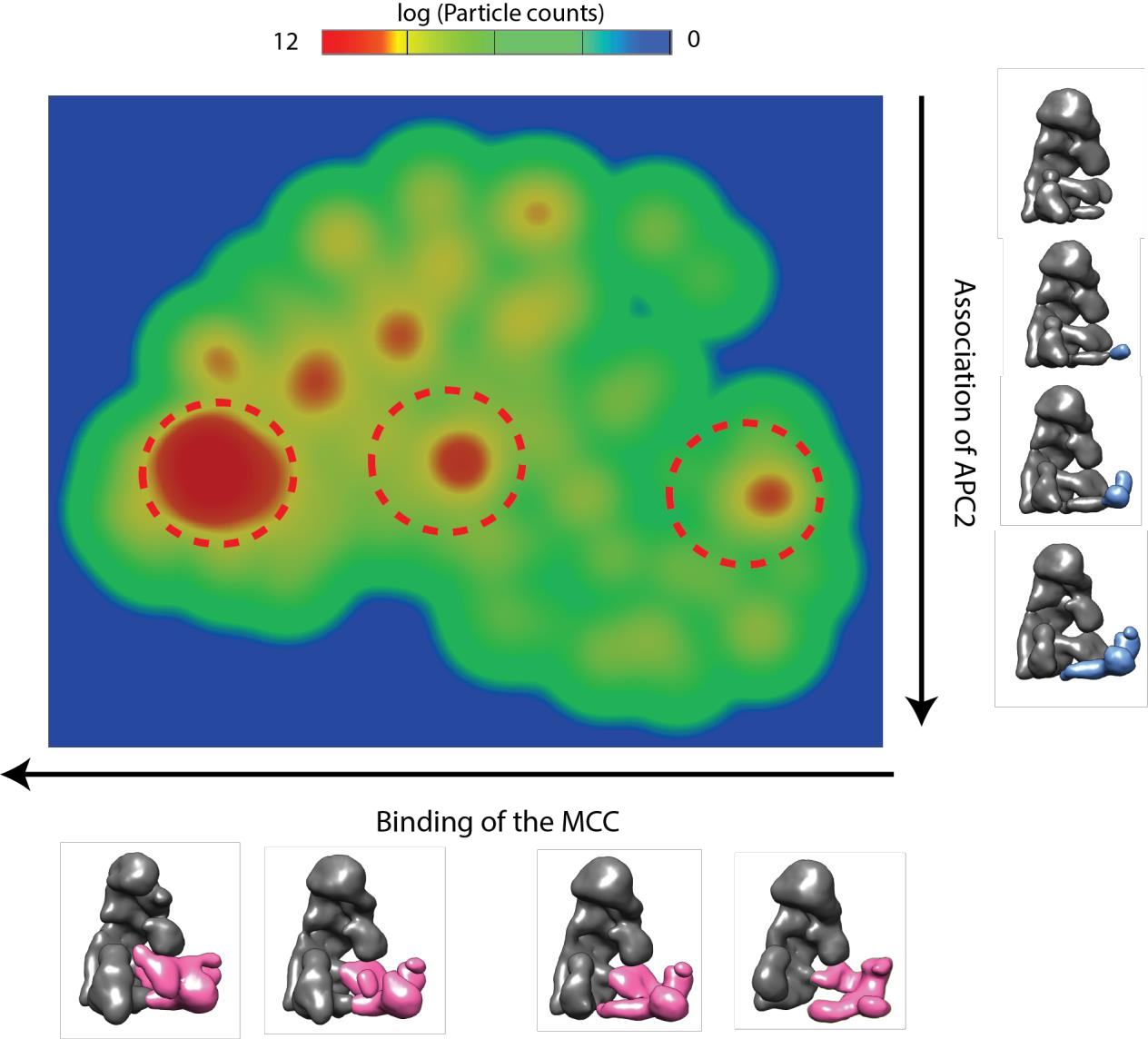}
  \caption{\textbf{Compositional landscape of APC.} This mainly compositional landscape visualizes the different assembly states of APC in a sample with a large degree of biochemical heterogeneity (i.e., it is not yet optimized in an early state of the project). The three major maxima correspond (red circles) to the different binding states of MCC (along the x-axis; shown in magenta). Any structural changes along this x-axis correspond to compositional heterogeneity of MCC binding as well as to some motion of the MCC towards the active center of the APC. The y-axis reveals the level of substoichiometric binding of the protein APC2 (shown in blue). The plot can be used as a quantitative tool to monitor and improve the biochemistry of complex purification and preparation. As such, it cannot be interpreted as an energy landscape, yet it still provides valuable information about the quality of the complex to be studied.}\label{fig:apc}
\end{figure}

3D classification provides a large number of structures in different conformations and the corresponding particle numbers that were used to calculate the structures.
However, at this stage, the set of 3D structures lacks any order relative to each other; instead, it simply reflects the conformation ensemble present in the dataset.
To obtain a quantitative description of the major structural similarities, all 3D structures are subjected to 3D-PCA.
3D-PCA yields the eigenvectors (3D volumes) that capture the structural variances in a hierarchical manner.
Typically, the first few eigenvolumes can be used to describe a large part of the heterogeneity in the dataset.
\cowscape{} uses this information to visualize continuous motions along the first
eigenvectors in a movie-like manner.
These animations can reveal the major modes of motion present in the dataset and the coupling of movable parts in the macromolecular complex.
The major eigenvectors may also serve as coordinate axes that span an energy landscape.
The energy landscape can be visualized as a heatmap, the colors indicate the relative Gibbs free-energy differences (Fig. \ref{fig:apc}).
All classified 3D structures can be mapped into the coordinate system (i.e. on top of the energy heatmap) that describes the major conformational variability.
The energy profile provides the user with valuable information about the number of conformational states and an estimate of the energy barriers separating them.

\cowscape{} is generally applicable as demonstrated by previous studies of the 26S proteasome \citep{Haselbach2017}, the human spliceosome \citep{Haselbach2018} and the fatty acid synthase \citep{Singh2020}.
We have also demonstrated its use on publicly available cryo-EM datasets
in the Electron Microscopy Public Image Archive (EMPIAR) database (Supplementary Methods).
These datasets also serve as additional examples of what can be learned from a \cowscape{} analysis.
All datasets were downloaded and processed in a similar manner (see Methods).
The 3D volumes were subjected to 3D-PCA, and the corresponding energy landscapes were calculated (Fig. \ref{fig:apc}) by Kernel Density Estimation with a Gaussian kernel function (Supplementary Methods).

\begin{figure}[h!]
  \centering\includegraphics[width=0.8\textwidth]{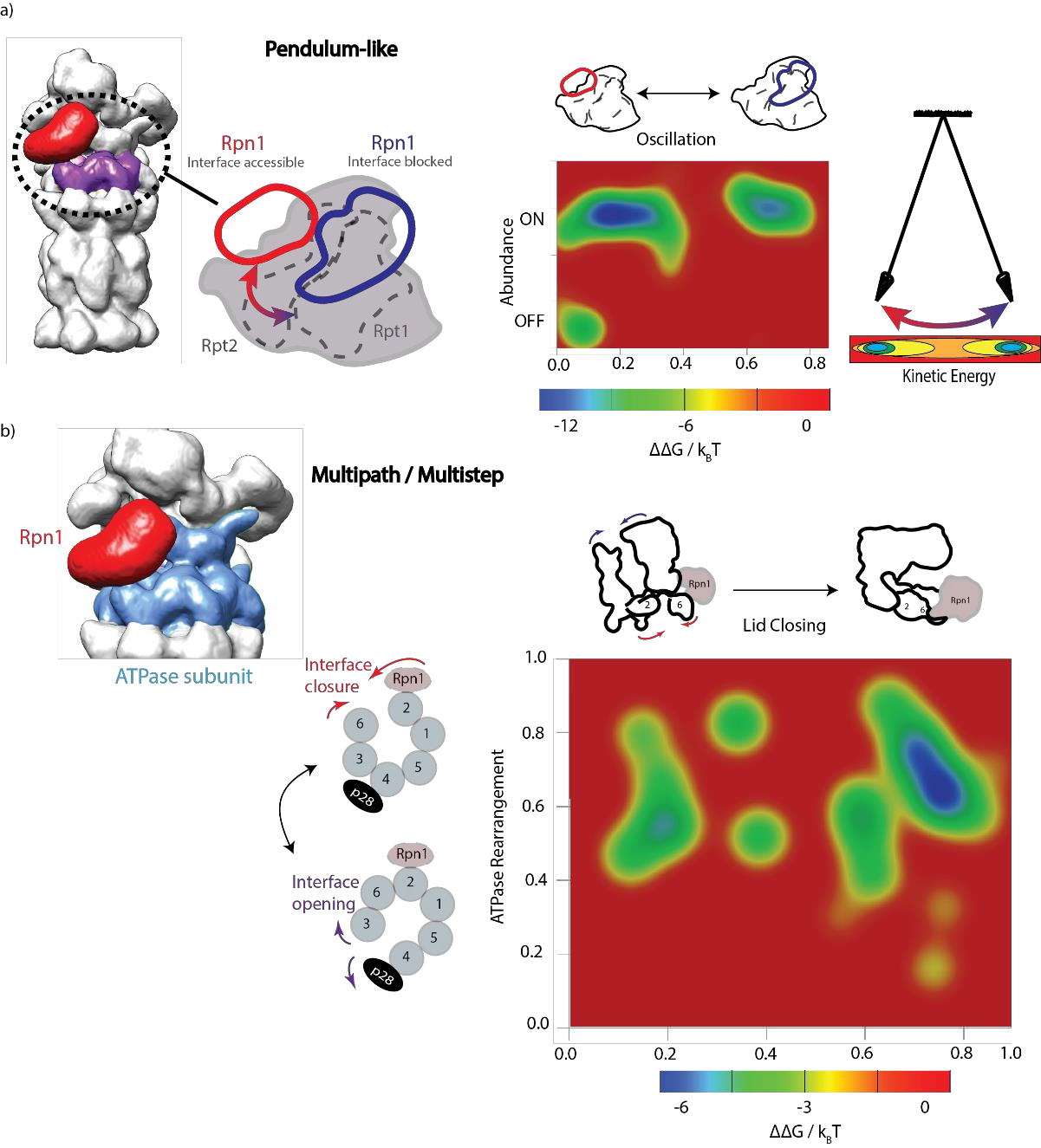}
  \caption{\textbf{Dynamic aspects within the 26S Proteasome.} a) RPN1 is a very dynamic protein that can adopt various conformations on the 26S proteasome. \cowscape{} analysis reveals the two major conformations of RPN1 with respect to the Rpt2/Rpt1 interface (purple) and a requirement for a pendulum-like motion in order to proceed from one major state to the other. The major variability along the x-axis of the energy landscape can be described by this pendulum-like movement (see also Supplementary Fig. 1) whereas the y-axis in the plot describes RPN1 abundance. The RPN1-ON state reveals an energy profile similar to a two-state energy profile of a pendulum, as schematically depicted. b) Detailed analysis of the dynamic behaviour of the ATPase subunit (blue) as a major component of the 19S regulatory subunit of the 26S proteasome. Whereas the major mode describes an opening and closing of the 19S subunit along the x-axis of the plot, the y-axis reveals a novel conformational change in the ATPase subunit that can be summarized as an opening and closing of the ATPase ring system via different subunit
interfaces.
  }\label{fig:proteasome}
\end{figure}

\begin{figure}[h!]
  \centering\includegraphics[width=0.8\textwidth]{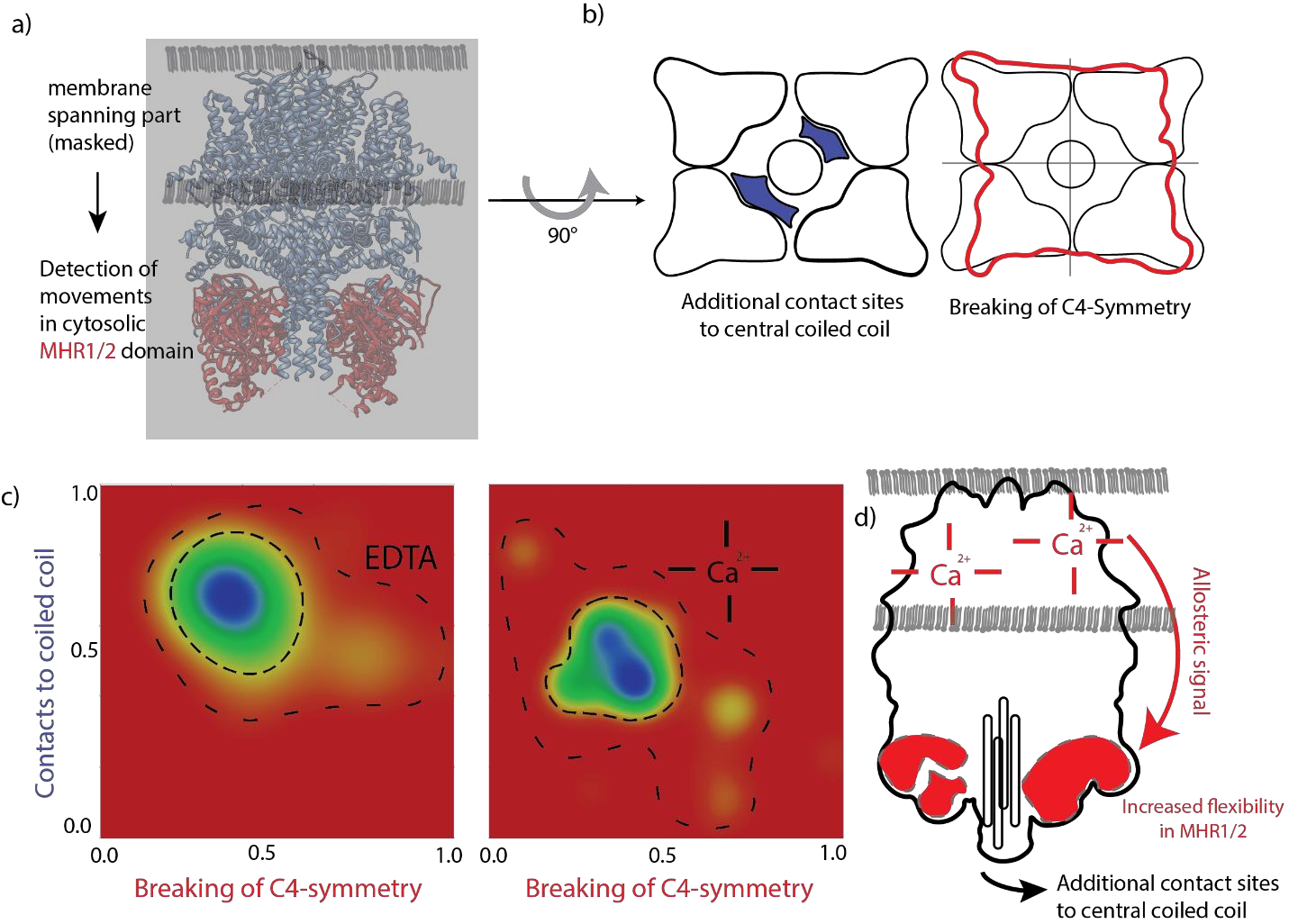}
  \caption{\textbf{Allosteric Ca$^{2+}$ signaling in the TRPM4 channel.} a) \cowscape{} analysis of the TRPM4 (transient receptor potential melastatin member 4) channel not only revealed the overall conformations 16 but also conformational variations that can be attributed to an overall change from a C4 to a (more-or-less) C2 symmetry behaviour (x-axis). Changes along the y-axis can be described by additional contact sites of the MHR1/2 domains being established to the central coiled-coil domain. b) Binding of calcium leads to an overall change in symmetry and to a conformational change that creates new internal contact sites. c) Even though the binding sites for calcium is on the membrane site, an increased flexibility of the MHR1/2 domains located at the opposite site can be observed, strongly suggesting that an allosteric signaling (d)
pathway affects the entire TRPM4 channel and finally leads to channel opening.
  }\label{fig:trpm4}
\end{figure}

Some general features of macromolecular complexes can be seen directly in the energy landscape plots (Fig. \ref{fig:proteasome}).
For some biochemically well-behaved complexes, such as the non-selective cation channel TRPM4 (Fig. \ref{fig:trpm4}), the energy landscape reveals well-defined thermodynamic minima (blue), suggesting that a large part of the data contributes to one major structure, and that there is only a relatively small number of distinct states in the dataset.
In contrast, other complexes such as the spliceosome reveal a continuously populated energy landscape, making it more difficult to find a sufficient number of images to determine structures of all conformational states at high resolution.
For these complexes, the energy barriers between the various states are much smaller, and the energy landscape can usually be modulated more easily by ligands that bind to the complex.

It is also noteworthy that, depending on the nature of the macromolecular complex, the energy landscape not only reflects the conformational heterogeneity of the complex, but can also contain sub-populations of complexes that differ in their composition.
\cowscape{} always analyzes the entire variability in the dataset, which comprises both conformational and compositional heterogeneity.
If the sample is known to be a very clean preparation, the population landscapes can be interpreted as differences in conformational sampling of the same complex, and thus assumed to correspond to an energy landscape.
In either case, the landscape provides the user with the unique possibility of discovering previously unknown conformations, for which one can subsequently try to select the raw data and refine them to high-resolution.
For instance, the landscape determined for the mitotic checkpoint complex (MCC) bound to the anaphase promoting complex (APC) \citep{Brown2016} revealed a high level of biochemical heterogeneity within the complex.
Therefore, Fig. \ref{fig:apc} cannot be interpreted as an energy landscape.
The two major eigenvectors of this dataset describe the stable integration of MCC into the APC complex and the presence or absence of the protein APC2.
Such a landscape can thus be used as a tool to quantitatively monitor any improvement in complex preparation and purification, with the goal of maximizing the number of molecules that integrate all desired components stably into the complex.
Energy landscapes can only be determined at a later stage after successful biochemical optimization. 

To focus on conformational variability, we analyzed examples that can be considered being biochemically optimized, namely, the 26S proteasome and the TRPM4 channel 16 (Figs. \ref{fig:proteasome} and \ref{fig:trpm4}).
For the 26S proteasome (Fig. \ref{fig:proteasome}), our analysis confirmed the macromolecular complex conformations that had been previously described \citep{Lu2017, Wang2017}.
However, the energy landscapes provided a more quantitative view about the data in
general that goes beyond previous findings.
Specifically, we applied \cowscape{} to two highly dynamic proteasome samples, focusing on the subunit RPN1 within the 26S proteasome holocomplex (Fig. \ref{fig:proteasome}a) and the regulatory 19S subcomplex bound to the chaperone p28 (Fig. \ref{fig:proteasome}b).
A focused classification and subsequent \cowscape{} analysis revealed an almost continuous pendulum-like motion for RPN1, which would explain why this protein has so far been elusive in high-resolution structure studies of the 26S proteasome.
The pendulum-like two-state distribution has its pivot point within the N-terminal coiled coil of RPT1 and RPT2 and might represent a regulatory mechanism by which the Rpt1/2 interface can be blocked (Supplementary Fig. \ref{fig:S1}).

In contrast to other observed motions, the mobility of the RPN1 pendulum does not seem to be directly coupled to the larger modes of motion that can be observed for the entire 26S
proteasome.
For the 19S proteasome, we observed opening of the Rpt2/Rpt6 interface in the ATPase ring structure with a simultaneous closing of the Rpt3/4 interface.
These changes correlate with a second interface closure at the RPN3 / RPN7 interface in the non-ATPase part of the complex (Supplementary Fig. \ref{fig:S2}).
Thus, using \cowscape{} we can not only automatically recover the order of minimum changes through the conformational snapshots but also directly distinguish between non-coupled and coupled motions within a macromolecular complex, which is likely to provide valuable information about the function of these large assemblies.

Figure \ref{fig:trpm4} shows another striking case for the TRPM4 channel with and without bound calcium ions (EMPIAR-10127 and EMPIAR-10126, respectively) \citep{Autzen2018}.
We detected an interesting conformational flexibility at the cytoplasmic side of the channel, which to our knowledge has not yet been described in the literature
(Fig. \ref{fig:trpm4}b, Supplementary Fig. \ref{fig:S3}).
By comparing the landscape with and without bound calcium, we can speculate that calcium binding increases the number of contact sites between the soluble channel parts
with the central coiled coil.
These soluble parts might be involved in second messenger signalling, and increased flexibility might hence precede channel opening.
The location of calcium binding is significantly distant from the detected conformational changes, which implies allosteric signalling.
The functional relevance of these changes in conformation sampling after binding small ligands has yet to be determined.

We previously showed for the 26S proteasome that the major conformation of the entire complex was almost identical irrespective of the presence or absence of the cancer drug oprozomib \citep{Haselbach2017}.
However, after drug binding to the 26S proteasome, the overall conformational space differed considerably \citep{Haselbach2017}.
A similar situation was observed for the TRPM4 channel.
Calcium binding to the TRPM4 channel is analogous to oprozomib binding to the 26S proteasome, with direct consequences on the ability of the large macromolecular complex to adopt the set of conformations that are relevant for its function. 
While the exact details of how this happens are still not known, it is possible to study such effects by calculating energy landscapes based on the analyses of all particle images present in a given dataset.
This also implies that \cowscape{} is a powerful tool for the structural interpretation of ligand binding to a macromolecular complex and for allowing functionally relevant conformational changes to be observed that would otherwise remain invisible (e.g., if they do not affect the stability of the major conformation).

The above examples illustrate that the quantitative analysis of conformational sampling is even more informative when two or more energy landscapes of the same complex but under different biochemical conditions can be compared.
Such comparative studies are ideally suited for elucidating how regulatory factors influence macromolecular machines in a more quantitative manner.
However, as analysis requires a minimal particle image frequency for a conformational state to be discovered, short-lived intermediates of a macromolecular complexes (such as those in transition states in catalysis) will not show up in this analysis, as they are not sufficiently populated.
How sensitive \cowscape{} can become at visualizing the conformational variability in macromolecular complexes in a quantitative manner will depend on the image statistics used for the \cowscape{} analysis and on the power of the applied 3D classification algorithms and the computational processing strategy.
While this may be a limitation at the moment, one can expect that detectors and computers become significantly faster in the near future, and that novel 3D classification algorithms can be developed that will be able to determine ever smaller
conformational differences.
The \cowscape{} algorithm itself is very fast and will not be a limiting factor. 

\section{Discussion}
\cowscape{} offers a comprehensive way to analyze and display ``motions'' in a cryo-EM dataset and to estimate quantitative free energy differences that can be used to deduce mechanisms underlying motions or allosteric signal propagation \citep{Bahar2010}.
%
%
\cowscape{} does not need any a priori information.
This makes it possible to use \cowscape{} at a very early stage of a project, thereby allowing the identification of the macromolecular complex conformations for which high-resolution structure refinement is best possible.
The kind of insight generated by \cowscape{} goes beyond the purely structural point of view.
It gains information that is comparable with, for instance, spectroscopic data, for which the distribution of states is usually accessed by only a one-dimensional output.
This powerful feature of the \cowscape{} analysis is strongly coupled to the idea of not discarding any true particle images from a given dataset.
In contrast, all the available data are used to obtain a quantitative understanding of the ensemble.
In the future, this may facilitate detailed quantitative experiments that describe how variables (e.g., temperature, pH values, salt conditions, or specific drugs) interfere with a macromolecular complex.
It will be difficult to obtain this information by any other method, yet such insight is critical for our understanding of how large macromolecular complexes act as ``molecular machines''.

\section*{Funding}
This work was funded by a grant of the Deutsche Forschungsgemeinschaft (DFG) within SFB 860 to HS.
MH, HS and AK are grateful for the support of the DFG within project 432680300 - SFB 1456 subproject A05. 
MH and AK gratefully acknowledge funding by the Carl Zeiss Foundation within the program ``CZS Stiftungsprofessuren''.

\clearpage

\section*{Figures}

\clearpage

\clearpage

\clearpage

\clearpage


\clearpage

\begin{appendix}

\renewcommand\thefigure{S\arabic{figure}}    
\setcounter{figure}{0}    

\renewcommand\thetable{S\arabic{table}}    
\setcounter{table}{0}    

\section*{Supplementary Figures}

\begin{figure}[h!]
  \centering\includegraphics[width=0.9\textwidth]{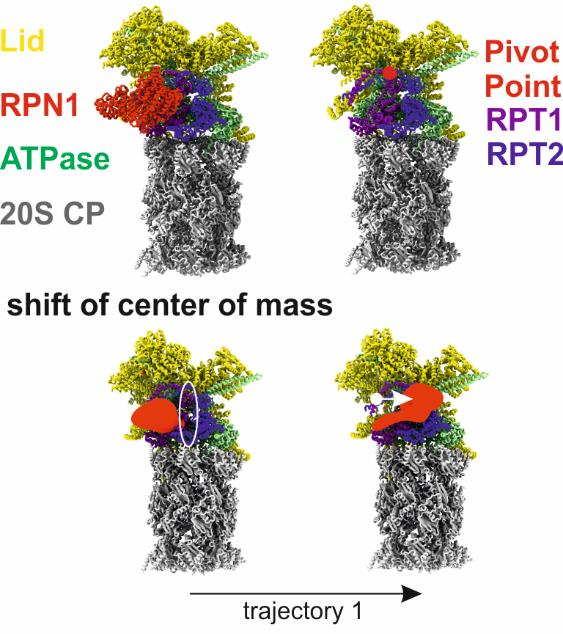}
  \caption{\textbf{RPN1 movement.}
    The upper panel shows the same view of the human 26S Proteasome (PDB 5M32) with and without the protein RPN1.
    RPN1 is labelled in red and the proteins RPT1 and RPT2 are coloured in purple and blue, respectively.
    \cowscape{} revealed a combined shift and pendulum-like movement of RPN1 described in the main text that reflects the major variability along the first eigenvector.
    RPN1's center of mass completely covers the RPT1/2 interface (white circle).
    RPN1 shifts from its position in front of RPT1 (white dot) towards RPT2 following the indicated movement vector.
  }\label{fig:S1}
\end{figure}

\clearpage

\begin{figure}[h!]
  \centering\includegraphics[width=0.9\textwidth]{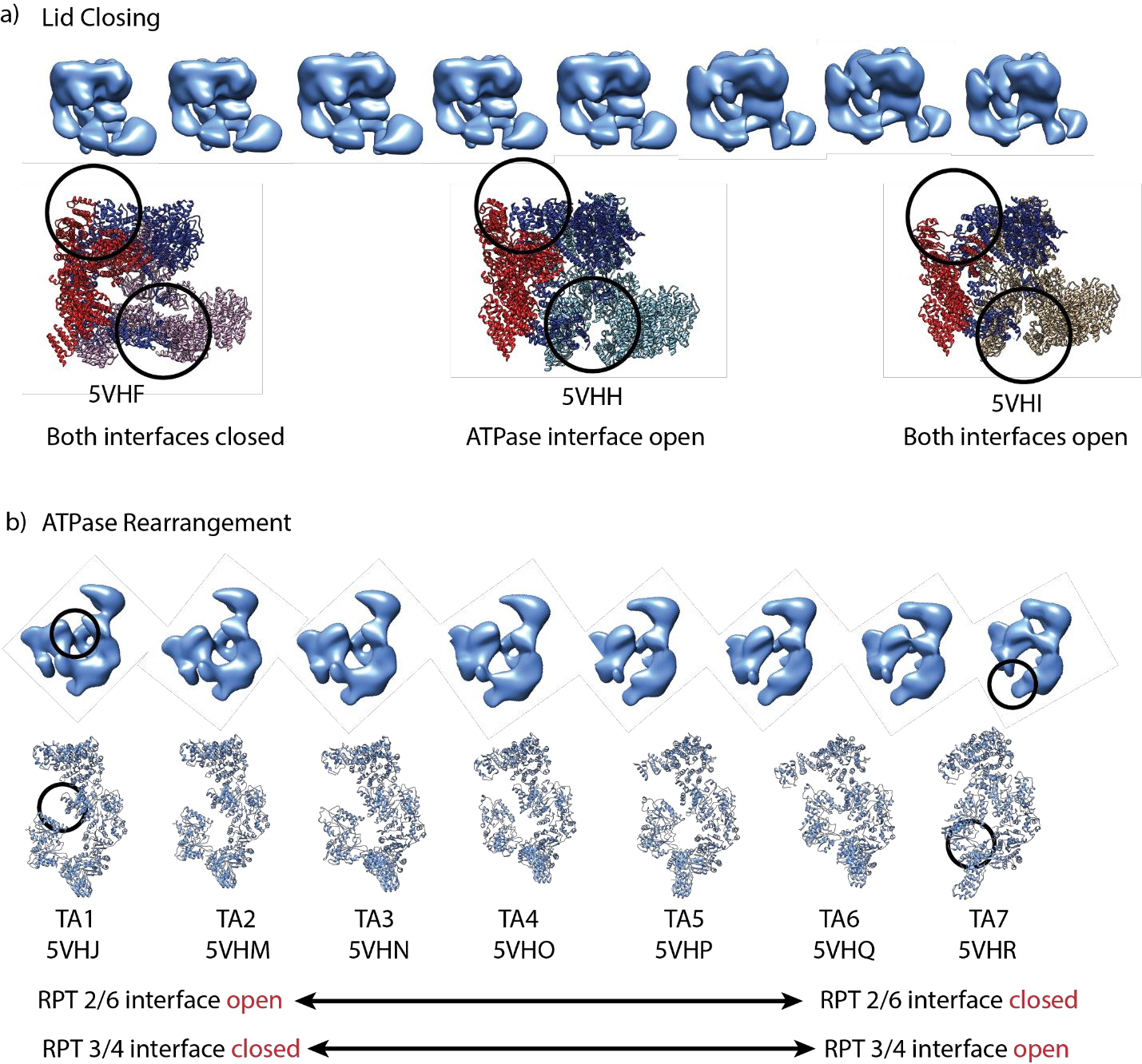}
  \caption{\textbf{p28-bound 19S proteasome.}
    Binding of p28 is a necessary component of the 26S assembly pathway, as it positions the 19S correctly on the 20S core particle, to form the full 26 proteasome.
    \cowscape{} detected -- automatically and without a priori information -- the findings described by \citet{Lu2017}.
    a) The results of the PCA analysis confirm the lid closing observed by Lu et al.
    b) Using \cowscape{} we determined another motion located in the ATPase ring system that correlates with the lid closing.
    Here we observe an alternating opening of the RPT2/6 and RPT3/4 interfaces.
    Based on the energy landscape shown in Fig. \ref{fig:proteasome}, we can conclude that the lid closes upon p28 binding.
    Simultaneously, the ATPase interface changes in going from an open RPT2/6 interface to an open RPT3/4 interface.
  }\label{fig:S2}
\end{figure}

\clearpage

\begin{figure}[h!]
  \centering\includegraphics[width=0.9\textwidth]{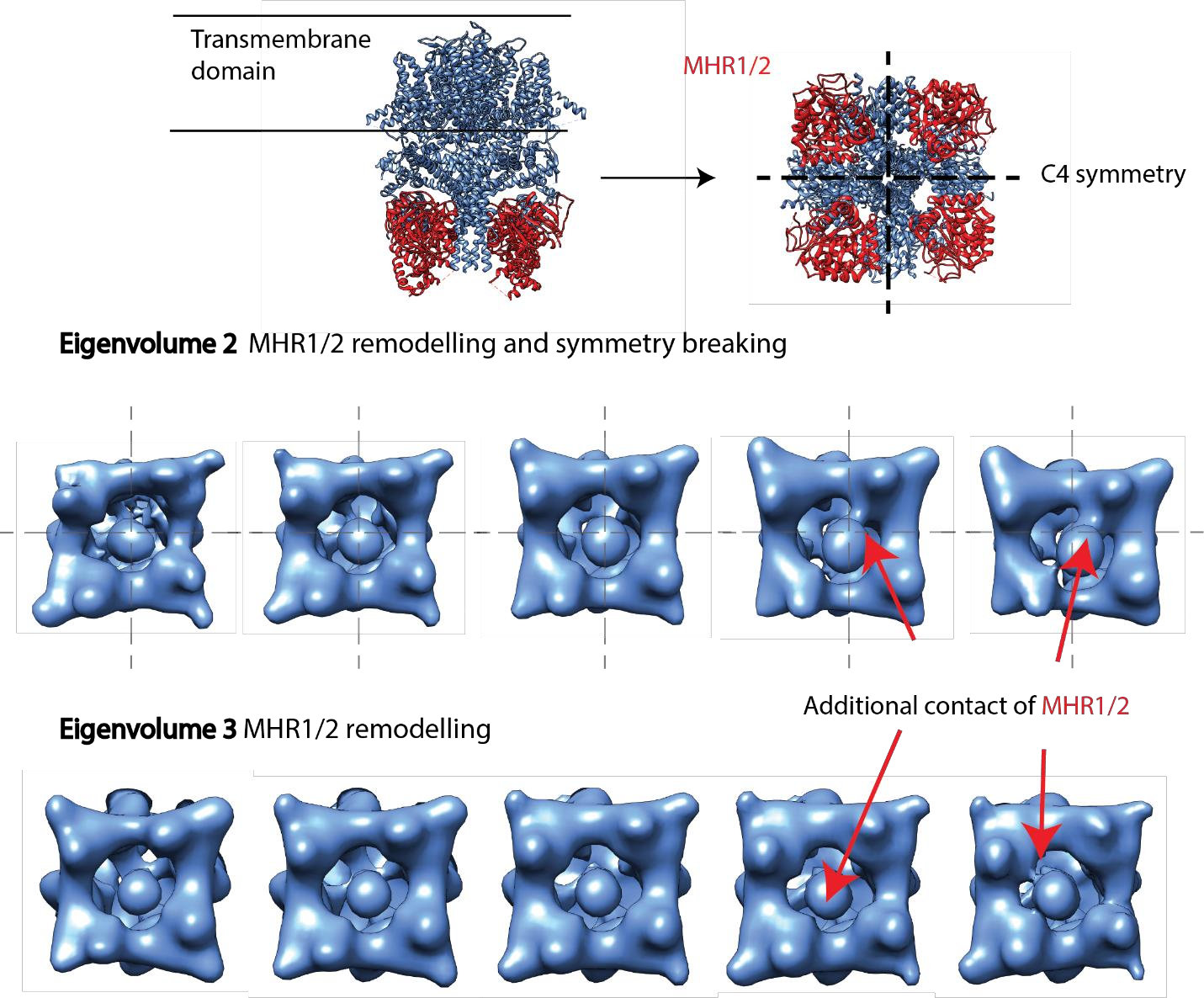}
  \caption{\textbf{Analysis of motions in the TRPM4 channel.}
    TRPM4 has been implicated in various severe diseases, such as diffuse large B-cell lymphomas and complete heart block.
    Our analysis focused on the cytoplasmic domain by applying a mask during PCA.
    Additionally, we did not restrict our analysis to C4 symmetry.
    The original structure (6BQV) is shown in the upper panel, with the MHR1/2
    domains marked in red.
    We detected large symmetry breaks in the second eigenvolume as well as additional contact sites of the MHR1/2 domains with the central coiled coil in the eigenvolumes 2 and 3.
    These contact sites were not observed in the original analysis by \citet{Autzen2018}.
    Interestingly, the MHR1/2 domains harbour putative binding sites for secondary messengers, such as PIP2 and PIP3.
    Flexibility seems to increase upon binding of calcium ions, which precedes channel opening.
    The observed additional binding sites to the coiled-coil might be the reason why Autzen et al. observed more side chains to be visible in the calcium bound state.
  }\label{fig:S3}
\end{figure}

\clearpage

\clearpage

\section{Data analysis}

All data were downloaded from the EMPIAR Database (Spliceosome: 10160, Proteasome: 10166, 19S particle with p28: 10091 and the TRPM4 channel 10126, 10127) and subjected to several rounds of 2D classification in the COW suite to remove non-particle views. As described in the main text, the general workflow divides into three main parts:
\begin{enumerate}
\item The generation of an ensemble of 3D structures.
\item The calculation and visual inspection of the reaction coordinates.
\item The construction of the energy landscape.
\end{enumerate}

\section{Ensemble Generation and Alignment}
We define an {\em ensemble} as a set of 3D volumes of the same size and voxel size.
Our approach can be applied to any conformational ensemble of 3D volumes.
Here, we describe an application to 3D volumes generated by extensive Maximum-Likelihood 3D classification, which is tested in RELION, SPHIRE and the COW Suite.
For all examples discussed in the manuscript, classes were prepared with classification runs containing 200,000 particles on 40 classes (if the initial dataset was smaller, fewer classes were used so as to match the same ratio between the number of particles and the number of classes).
It is necessary to split the entire population of particles into multiple batches which are classified separately, because the computational effort increases dramatically with the number of particles.
After classification, individual volumes where autorefined and all volumes were pooled into a single ensemble.
Because PCA is sensitive to any difference between the volumes, aligning them against the same reference is crucial.
This alignment was carried out either in the COW Suite using the Logic \textit{3DAlignment} with standard parameters or in UCSF Chimera \cite{} with the \textit{FitinMap} procedure.
3DAlignment maximizes the real space cross-correlation coefficient to a reference by applying shifts and rotations to the volume.
Some remarks regarding the specifics of processing the data sets are listed below: \\

\begin{center}
  \begin{tabular}{l l}
    \toprule
    \textbf{Dataset} & \textbf{Remarks on classification and processing} \\
    \midrule
    \multirow{3}{*}{26S/rpn1} & \multirow{3}{*}{\parbox{0.73\textwidth}{
        No alignment was necessary, because the classification was only local and the 20S part remained stable throughout the classification runs.}} \\[1cm]
    \multirow{3}{*}{TRPM4}  & \multirow{3}{*}{\parbox{0.73\textwidth}{
        The classification and subsequent refinement was performed without applying symmetry restraints (in contrast to the original publication).}}\\[1cm]
    \bottomrule
  \end{tabular}
\end{center}

All subsequent steps were performed within \textit{CowEyes}.
The volumes were lowpass-filtered to the lowest resolution among all ensemble members.
This is necessary to avoid that PCA picks up resolution differences between the volumes.
The Fourier space cutoff of the filter is given by 
\[
\gamma = \frac{2 \times{} \text{pixelSize}}{\text{lowest resolution}}\\[0.1cm]
\]
Subsequently, all volumes were normalized to a mean of zero and a standard deviation $\sigma$ of 2.
The parameter $\sigma$ is an important variable to fine-tune the information content of the eigenvolumes, as it directly influences the detection of information by PCA in the next step.

\section{Principal Component Analysis (PCA)}
PCA finds an embedding of the input data in a lower dimensional space that preserves the correlation structure of the data as much as possible (for a thorough review of PCA see \citet{VanHeel2016}).
The coordinate axes of the lower dimensional space are obtained by calculating the eigenvectors of the covariance matrix of the input data (the ensemble of volumes). 
The eigenvectors are orthogonal to each other and sorted by the amount of covariance that each of them represents (their corresponding eigenvalue).
CowEyes' implementation of PCA is memory efficient especially for large volumes, since it uses the dimensional $N \times N$ covariance matrix where $N$ is the number of volumes.
By selecting a subset of the eigenvectors we construct a lower dimensional space that represents the input.
The position of the volume in this lower-dimensional space is determined by a vector of linear factors $a$ where $a_n$ (representing the $n$-th volume) has as many entries as the dimensionality of the embedding space.
COW calculates the linear factors by a simple matrix multiplication performed in the Logic \textit{PCATransform}.

\section{Visual Inspection of the Eigenvectors}
To decide which conformational transitions are captured by the eigenvectors, one needs to visually inspect them.
Therefore, trajectories are simulated along one eigenvector.
The algorithm for the generation of trajectories is implemented as \textit{TrajectoryVisualizer} in CowEyes.
It shows the information represented by a single user-specified eigenvolume by simulating a trajectory of $N$ volumes.
From the linear factors calculated in \textit{PCATransform} one could retain each volume $V_n$ in the dataset through the relation
\begin{equation}\label{eq:pca}
  V_n = \overline{V} + \sum_{k=1}^K a_{nk} E_k
\end{equation}
where $\overline{V}$ is the average volume calculated in the PCA, $V_n$ is the $n$-th volume in the ensemble, $E_k$ is the $k$-th eigenvolume describing a conformational motion and $a_{nk}$ are the linear factors that act as coordinates of $V_n$ in the space spanned by $E_k$.

\textit{TrajectoryVisualizer} simulates a series of volumes that only displays the information encoded by a single eigenvolume $E_k$.
Hence Eq. \ref{eq:pca} simplifies to
\begin{equation}\label{eq:animate}
  V_k = \overline{V} + a_k \, E_k\, .
\end{equation}
To show the $N$ snapshots along the trajectory, first the linear factors of all volumes with respect to the specified eigenvector $E_k$  are collected and sorted into one vector $a_k = (a_{1k}, \ldots, a_{Nk})^T$, where $a_{1k}$ and $a_{Nk}$ are the smallest and largest linear factors, respectively.
The number of principal components is $K \le N$, because typically the size of the volume (i.e. the number of voxels) is larger than the size of the ensemble (i.e. the number of volumes). 
The average volume $\overline{V}$ is constant throughout the dataset.
\textit{TrajectoryVisualizer} now offers two modes:
Either the real linear factors are used for the calculation (i.e. all elements of $a_k$) or an equally sampled interval between $a_{1k}$ and $a_{Nk}$ is used.
In the latter case, a trajectory sampled with a stepsize of $(a_{Nk} - a_{1k}) / (N-1)$ is visualized.
By applying Eq. (\ref{eq:animate}) for each element in the vector $a_k$, or for each sample in the interval $[a_{1k}, a_{Nk}]$, respectively, a volume series is calculated.
The snapshot volumes along the trajectory can be visualized in the Volume viewer of CowEyes.
Supplementary Videos were generated by loading the full trajectory into UCSF Chimera and recording a movie of a morph between the snapshots (a Chimera script is provided in the supplement).

\section{Energy Landscape Estimation}
The previous analysis steps mainly serve us to find an efficient representation of the input 3D volumes in a low-dimensional space.
What is still missing is a continuous description of this space that also allows the estimation of energy differences between ensemble members.
The algorithm for the estimation of a continuous surface from the linear factor representation is also implemented in CowEyes as \textit{EnergyLandscape}.
From two chosen eigenvolumes $E_1$ and $E_2$, the linear factors are saved in two vectors $a_1 = (a_{11}, \ldots, a_{N1})^T$ and $a_2 = (a_{12}, \ldots, a_{N2})^T$, respectively.
Additionally, a third vector $c = (c_1, \ldots, c_N)^T$ containing the observation frequencies or particle counts of the corresponding volume is stored.
Each data point of the form $(a_{n1}, a_{n2}, c_n)$ is then mapped to a grid forming a weighted point cloud.
To estimate a continuous surface, kernel density estimation (KDE) with a Gaussian
kernel function 
\[
\phi_n(x, y) = c_n e^{-\frac{1}{2\sigma^2} ((x - a_{n1})^2 + (y - a_{n2})^2)}
\]
is used where $x$ and $y$ are the centers of the cells of the sampling grid.
The Gaussian kernels are evaluated at each grid point $(x, y)$ and added together resulting in a continuous probability surface:
\begin{equation}\label{eq:kde}
\phi(x, y) = \sum_n \phi_n(x, y) = \sum_n c_n e^{-\frac{1}{2\sigma^2} ((x - a_{n1})^2 + (y - a_{n2})^2)}
\end{equation}
The bandwidth $\sigma$ determines the width of the Gaussian and has to be user-adjusted.
With the standard grid spacing, all landscapes shown are plotted with a $\sigma$ of 300px.
An energy landscape is obtained by computing $-\log \phi(x, y)$, which follows from Boltzmann inversion.
Alternatively, we can use Gaussian kernel regression to approximate the energy landscape directly by relacing the counts in (Eq. \ref{eq:kde}) with temperature-normalized Gibb's free energies:
\[
\left(\frac{\Delta G}{k_B T}\right)_n = \ln\left(\frac{c_n}{\max \{c_n\}}\right)
\]
A final 3D-rendered visualization of the landscape can be obtained by CowEyes' built-in surface viewer, which is based on the Qt Data Visualization library.

\hide{

\section{Further Discussion}
We have outlined the algorithmic workflow to analyze an ensemble of cryo-EM volumes with respect to their intrinsic modes of motion.
These motions are used to describe the functional region of the energy landscape that the macromolecular machine visits in solution.
Our approach is based on competitive Maximum Likelihood Refinement (``3D classification'') which generates the volumetric ensembles used in this work.
3D classification assigns one particle to a randomly initialized reference. 
However, the generation of this ensemble is extremely flexible and not restricted to 3D.
Alternatively, an ensemble of 2D images representing the same view could also be used.
This approach would not require the 3D classification step, which is computationally expensive.
A 3D analysis necessitates that a certain subset of 2D images has been averaged during 3D reconstruction.
A drawback of this approach is that it discretizes the continuous movement of macromolecular machines into those volumes forming the ensemble.
However, averaging 2D images suppresses noise and hence increases the robustness and information content of the statistical analysis by PCA.
The last point is a clear advantage over statistical methods based purely on 2D images.

\hide{PCA was used before to detect conformational motions in cryo-EM data on both 2D \citep{White2004} and 3D data \citep{Fischer2010, Klaholz2004, Penczek2006}.
However, PCA only finds a linear approximation to the motions in the dataset.
This also implies that the real motions of the molecule are always a composition of all eigenvolumes found for the dataset.
Non-linear approaches are in contrast prone to overfitting and their validity has to be verified thoroughly in the future.
Moreover, \cowscape{} uses the eigenvolumes primarily for dimensionality reduction of the high-dimensional volumes.
For this purpose, PCA offers a robust way of finding dimensions representing conformational motions \cite{Penczek2006} and its application to volumes rather than coarse models increases the information content further.
}

After embedding the volumes into the low-dimensional space, \cowscape{} quantitatively interprets the experiment by deriving Gibb's free energies.
The free energies are estimated from the number of particles per 3D class by Boltzmann inversion.
Counting information resulting from 3D classification algorithms has been used only rarely.
The estimated energies are prone to errors for the following reasons:
i) Broken or empty particle images are more likely assigned to abundant conformations, because these contain more high-resolution information with which the noise in the images correlates.
Hence, these images overestimate energy differences.
ii) Volumes showing similar conformations can exhibit different counts for several reasons. 
The refinement step after 3D classification involves a hard assignment to the volumes, thereby introducing discontinuous steps in the energy landscape.
The last argument is addressed by estimating energy surfaces using Gaussian KDE.
KDE will a-posteriori soften the assignments in the 3D classification and increase the robustness of the energies.
A Gaussian kernel has two parameters: its mean $\mu$ and its standard deviation or bandwidth $\sigma$.
Since the means are fixed by the position in 2D space, the bandwidth is critical for the absolute height of the energies and still has to be adjusted.
In \cowscape{} this is left to the expertise of the user (Supplementary figure).
Future improvements of the model have to include a stable estimation procedure of the bandwidth parameter. 

The computationally most expensive step is the generation of the conformational ensemble by 3D classification.
In comparison, the \cowscape{} analysis involves a negligible computational effort. 
Hence, the analysis and testing of different parameters can be done just in time after the ensemble was generated once.  

}

\end{appendix}

\end{document}